\documentclass[letterpaper,12pt]{article}
\usepackage{amssymb}

\usepackage{ifpdf}
\ifpdf
	\usepackage[pdftex]{graphicx}
	\usepackage[pdftex,unicode,implicit]{hyperref}

	\hypersetup{
  	pdftitle     = {Consistency of the Hamiltonian formulation of the lowest-order effective action of the complete Ho\v{r}ava theory},
  	pdfkeywords  = {},
  	pdfauthor    = {J. Bellor\'{\i}n and A. Restuccia},
  	pdfcreator   = {pdf\LaTeXe\ with package \flqq hyperref\frqq},
  	pdfproducer  = {pdf\LaTeXe\ with package \flqq hyperref\frqq},
  	pdfpagemode  = UseNone,  
  	pdffitwindow = true,  
  	unicode      = true,
  	plainpages   = true,
  	colorlinks   = true,  
  	citecolor    = blue,  
  	urlcolor     = blue,
  	linkcolor    = blue
	}

\else
  \usepackage[dvips]{graphicx}
  \usepackage[unicode,implicit]{hyperref}

\fi

\makeatletter
\@addtoreset{equation}{section}
\makeatother

\begin{document}

\begin{flushright}
{\small
Dec $7^{\rm th}$, $2011$}
\end{flushright}

\begin{center}

\vspace*{2cm}
{\bf \LARGE 
Consistency of the Hamiltonian formulation of the lowest-order effective action of the complete Ho\v{r}ava theory} 
\vspace*{2cm}

{\sl\large Jorge Bellor\'{\i}n}$^{a,}$\footnote{\tt jorgebellorin@usb.ve}
{\sl\large and Alvaro Restuccia}$^{a,b,}$\footnote{\tt arestu@usb.ve}
\vspace{3ex}

$^a${\it Departamento de F\'{\i}sica, Universidad Sim\'on Bol\'{\i}var, Valle de Sartenejas,\\ 
1080-A Caracas, Venezuela.} \\[1ex]
$^b${\it Department of Physics, Universidad de Antofagasta, Chile.}

\vspace*{2cm}
{\bf Abstract}
\begin{quotation}{\small
We perform the Hamiltonian analysis for the lowest-order effective action, up to second order in derivatives, of the complete Ho\v{r}ava theory. The model includes the invariant terms that depend on $\partial_i \ln N$ proposed by Blas, Pujol\`as and Sibiryakov. We show that the algebra of constraints closes. The ``Hamiltonian'' constraint is of second-class behavior and it can be regarded as an elliptic partial differential equation for $N$. The linearized version of this equation is a Poisson equation for $N$ that can be solved consistently. The preservation in time of the Hamiltonian constraint yields an equation that can be consistently solved for a Lagrange multiplier of the theory. The model has six propagating degrees of freedom in the phase space, corresponding to three even physical modes. When compared with the $\lambda R$ model studied by us in a previous paper, it lacks two second-class constraints, which leads to the extra even mode.
}\end{quotation}

\end{center}

\thispagestyle{empty}

\newpage
\section{Introduction}
There has been a debate about the consistency of the Ho\v{r}ava theory \cite{Horava:2009uw}, which is a proposal for a UV completion of general relativity that could be a renormalizable theory. Some of the papers involved in this debate can be found from Refs.~\cite{Kobakhidze:2009zr} to \cite{varios:perturbativeblasmodel}. The discussion has been mainly focused on the presence of a physical mode additional to the ones of general relativity that could be the potential source of instabilities. This extra mode is present at all scales in the projectable version of the theory since in this case there is no local Hamiltonian constraint (see \cite{Kobakhidze:2009zr} for the Hamiltonian analysis on the corresponding second-order action)\footnote{In the projectable case the extra mode could be eliminated by enhancing the gauge symmetry group, see \cite{Horava:2010zj}}. On the other hand, we found \cite{Bellorin:2010je} that the lowest-order truncation of the original Ho\v{r}ava theory is physically equivalent to general relativity for arbitrary values of the constant $\lambda$. This result encourages us to deepen the study on the nonprojectable formulation of the Ho\v{r}ava theory, to which this paper is devoted.

Among the papers focused on the nonprojectable case, in Ref.~\cite{Charmousis:2009tc} perturbative computations signaled the problem of the strong coupling of the theory (in particular of the extra mode), which implies the breakdown of its low-energy perturbative expansion. This behavior was confirmed in Ref.~\cite{Blas:2009yd}. However, the extra mode did not manifest itself (as a propagating mode) in the linear-order perturbative analyses of Refs.~\cite{varios:linearperturbationsdeformed,varios:linearperturbations}\footnote{In Ref.~\cite{varios:linearperturbationsdeformed} some ''deformed'' versions of the theory were analyzed; but such deformations, although representing violations of the detailed balance principle, are just redefinitions of some constants in the potential. We prefer to keep their denomination as the ``original'' Ho\v{r}ava theory, in the sense that their potentials are written in terms of only the spatial curvature tensors (and the cosmological constant). In general we consider the several coupling constants as independent parameters; hence we do not use the detailed balance principle.}. This apparent contradiction is explained by the argument given in \cite{Blas:2009yd}, where it was indicated that the extra mode is excited in perturbative analyses (at linear order) only on time-dependent and spatially nonhomogeneous backgrounds\footnote{In addition, in Ref.~\cite{varios:debateproyectable} the physical behavior of the extra mode in the projectable formulation of the theory has been analyzed and other modifications/extensions have also been considered (see, for example, \cite{varios.debateothermodifications}).}. 

In Refs.~\cite{Li:2009bg,Kocharyan:2009te,Henneaux:2009zb,Pons:2010ke,Bellorin:2010je,Bellorin:2010te} the constraints of the original nonprojectable theory have been studied, as well as its lower-derivative truncations. The authors of Ref.~\cite{Li:2009bg} met several difficulties when performing an early Hamiltonian analysis. They could not close the algebra of constraints and suggested the arising of an excessive number of constraints (a result that has some similarity with \cite{Pons:2010ke}). Actually, these inconveniences are circumvented once the Hamiltonian constraint is regarded as a second-class constraint with no associated gauge symmetry \cite{Henneaux:2009zb,Bellorin:2010je,Bellorin:2010te}. Nevertheless, the authors of Ref.~\cite{Li:2009bg} noticed the fact that the phase space of the original Ho\v{r}ava is of odd dimensionality. In \cite{Kocharyan:2009te} it was proposed that the nonprojectable theory is in general inconsistent due to the presence of a constraint that is absent in general relativity. However, for the models studied in Refs.~\cite{Bellorin:2010je,Bellorin:2010te} this constraint can be solved in closed ways. Still on Hamiltonian grounds, in Ref.~\cite{Henneaux:2009zb}, assuming standard asymptotic behavior for the gravitational variables (the same assumptions we are going to use in this paper), it was argued that the lapse function must be vanishing at least asymptotically in the general theory and everywhere for the lowest-order truncation. The same posture was assumed in Ref.~\cite{Pons:2010ke}. Regarding this, in Refs.~\cite{Bellorin:2010je,Bellorin:2010te} we performed Hamiltonian analysis for low-energy effective models for the Ho\v{r}ava theory, starting in \cite{Bellorin:2010je} with the lowest-order model,
\begin{equation}
 S = 
 \int dt d^3 x \sqrt{g} N ( K_{ij} K^{ij} - \lambda K^2 + R ) \,.
\label{lambdar}
\end{equation}
As we mentioned above, in that paper we found that the model (\ref{lambdar}) is physically equivalent to general relativity regardless of the value of the constant $\lambda$, including the fact that it propagates two physical modes. This is due to the fact that, when performing Dirac's algorithm for the preservation of constraints, the condition $K = 0$ emerges as a second-class constraint of the model (instead of having a vanishing lapse function, as was proposed in \cite{Henneaux:2009zb}). Thus, the term $\lambda K^2$ drops out and the theory resulting from (\ref{lambdar}) evaluated on the constrained phase space coincides with general relativity in the particular gauge $K=0$. Next, in Ref.~\cite{Bellorin:2010te} we incorporated an $R^2$ term in the potential. We found that the algebra of constraints closes, but some peculiarities arise. First, the $R^2$ model lacks one of the second-class constraints of the model (\ref{lambdar})\footnote{When one considers $N$ and its conjugated momentum as part of the canonical variables, the constraints of the model (\ref{lambdar}) are the momentum constraint and four second-class constraints: the vanishing of the momentum of $N$, the Hamiltonian constraint, the vanishing of $K$, and an elliptic equation for $N$. The last three are the (ordered) chain of constraints yielded by imposing the preservation in time of previous constraints. These constraints reduce from 14 to four the number of canonical variables.}, which leads to a physical odd-dimensional (five dimensions) submanifold in phase space, as was previously indicated in Refs.~\cite{Li:2009bg,Blas:2009yd}. This means that there is a physical mode with a first-order evolution equation, hence propagating half of the usual Cauchy data. It seems that this half mode will be persistent when adding higher-order ($z=3$) terms in the potential. This is so because the lacking of one second-class constraint of (\ref{lambdar}) is a consequence of the explicit dependence on the lapse function of the constraint that would generate it. Whenever a (second-class) constraint depends on the lapse function, it is to be expected that its preservation in time generates an equation for a Lagrange multiplier of the theory, rather than a new constraint. This feature does not arise in (\ref{lambdar}), but probably will hold in general in the \emph{original} Ho\v{r}ava formulation with higher-order terms. Second, $N$ is determined by a first-order partial differential equation (PDE) of flow type. The boundary data compatible with such an operator must be given on a noncharacteristic surface defined by a certain vector field. In any case, the equation is compatible with the expected asymptotic behavior of the variables that yields the Minkowski space at infinity. Moreover, in the perturbative analysis around a weakly varying background we did in Ref.~\cite{Bellorin:2010te}, we found that the extra mode decouples at the lowest orders, hence recovering the physics of the model (\ref{lambdar}) smoothly. The results of Refs.~\cite{Bellorin:2010je,Bellorin:2010te} were corroborated by the perturbative Hamiltonian analysis of Ref.~\cite{Das:2011tx}.

As we mentioned above, the presence of the extra mode in Ho\v{r}ava theory was analyzed in Ref.~\cite{Blas:2009yd}. Those authors first made a counting of degrees of freedom in a model having up to a fourth-order term ($R_{ij} R^{ij}$) in the potential. The counting was formulated as a problem of initial Cauchy data in the corresponding equations of motion. Second, they focused on the physical behavior of the extra mode, but truncating to a second-order model with the aim of simplifying the computations. From their analysis they concluded that the extra mode suffers from very fast instabilities and strong coupling at the deep IR. However, it turns out that the second-order model they used to analyze the propagation of the extra mode is exactly the model (\ref{lambdar}), which does not contain the extra mode (this was already pointed out by us in Ref.~\cite{Bellorin:2010te}). Based upon this and the smooth decoupling found in \cite{Bellorin:2010te}, we suggest being cautious with the behavior of the extra mode of the nonprojectable Ho\v{r}ava theory.

The authors of \cite{Blas:2009yd} suggested that the inconveniences they found with the extra mode could be cured by the inclusion of additional terms depending on the spatial vector $a_i \equiv \partial_i \ln N$. These terms were not considered by Ho\v{r}ava in his original formulation. The proposal was formally presented by them in Ref.~\cite{Blas:2009qj}. Several terms that are invariant under the action of foliation-preserving diffeomorphisms, which are the underlying gauge symmetries of the Ho\v{r}ava theory, can be formed with appropriate combinations of $a_i$, its derivatives and curvature tensors. In particular, the expressions $\nabla_i a^i$ and $a_i a^i$ are second-order terms that are invariant under the foliation-preserving diffeomorphisms.

The physical consistency of the Ho\v{r}ava theory with the terms of Blas, Pujol\`as and Sibiryakov has also been the subject of study. The problem of the strong coupling was reported to persist in Ref.~\cite{Papazoglou:2009fj}. However, a response from Blas, Pujol\`as and Sibiryakov \cite{Blas:2009ck} (see also \cite{Blas:2010hb}) pointed out that the complete theory can be made free from the strong coupling problem if the energy scale of its UV physics (where higher-derivative terms become important) is below the Planck scale. In Ref.~\cite{Kimpton:2010xi} it was confirmed that the theory exhibits strong coupling if no new scales are introduced. Perturbative analyses of this theory with cosmological interest have been made in Ref.~\cite{varios:perturbativeblasmodel}.

Independently of the presence or not of undesirable physics behavior for the extra mode, specially taking into account that it decouples smoothly at the largest distances in the original Ho\v{r}ava model \cite{Bellorin:2010te}, one may be in favor of including the $a_i$ terms by following an argument of completeness of the theory. When dealing with effective theories it is normally assumed that all the terms of the order in consideration that are compatible with the gauge symmetries must be included in the Lagrangian. Therefore, if the quantum consistency of the Ho\v{r}ava could be proved, then the $a_i$ terms should be included since they will arise in quantum corrections\footnote{The renormalizability of the Ho\v{r}ava theory has been studied in Ref.~\cite{varios:renormalizacion}}.

With the aim of gaining more insight into the canonical structure of the nonprojectable Ho\v{r}ava theory, in this paper we study the Hamiltonian formulation of the lowest-order effective action (second-order in derivatives) that includes the $\nabla_i a^i$ and $a_i a^i$ terms. A previous Hamiltonian analysis of the nonprojectable Ho\v{r}ava theory with the terms of Blas, Pujol\`as and Sibiryakov was presented in Ref.~\cite{Kluson:2010nf}. In this paper we further develop the analysis of the constraints. In Sec.~II we shall find all the constraints of the theory explicitly and we shall see that Dirac's algorithm for extracting the constraints of the theory ends with an elliptic PDE for a Lagrange multiplier of the theory. This closure follows the same behavior we have mentioned: the last constraint being generated is the one that depends on the lapse function. This analysis will allows us to count the number of physical degrees of freedom of the model. In Sec.~III we shall show an interesting canonical transformation that is defined in terms of a conformal transformation and that leads to some simplifications in the canonical action.


\section{Analysis of the constraints}
The theory is written in terms of the Arnowitt-Deser-Misner variables
\begin{equation}
ds_4^2 = ( - N^2 + N_i N^i ) dt^2 + 2 N_i dx^i dt + g_{ij} dx^i dx^j \,.
\end{equation}
We denote by $\pi^{ij}$ the momentum conjugated to $g_{ij}$ and by $\phi$ the one of $N$. After performing the Legendre transformation it turns out that the vector $N_i$ can be regarded as a Lagrange multiplier, which is to be expected because of the gauge symmetries of the theory. This is not the case for $N$, which, together with its conjugated momentum, must be considered as part of the canonical variables. Hence the unconstrained phase space is parametrized by $\{ (g_{ij},\pi^{ij}),(N,\phi)\}$. $\pi^{ij}$ and $\phi$ behave as scalar densities under spatial coordinate transformations. 

We want to study the model as a local theory of gravity without topological effects. Therefore we assume that the whole spatial manifold is noncompact. In our variational calculus all the configurations are fixed at spatial infinity and, since we do not include a cosmological constant, correspond to Minkowski space-time. We also assume that the field variables have the same asymptotic behavior of general relativity \cite{Regge:1974zd}: in asymptotically flat coordinates, as $r$ approaches to infinity, they behave as
\begin{equation}
\begin{array}{rclrcl}
 g_{ij} & = & \delta_{ij} + \mathcal{O}(r^{-1}) \,, 
\hspace{2em}
 & \pi^{ij} & = & \mathcal{O}(r^{-2}) \,,
\\[2ex]
 N & = & 1 + \mathcal{O}(r^{-1}) \,, 
 & N_i & = & \mathcal{O}(r^{-1}) \,.
\end{array}
\label{asymptotics}
\end{equation}
Note that for the consistency of the canonical action it is assumed that $\dot{g}_{ij}$ and $\dot{N}$ go as $\mathcal{O}(r^{-2})$. Indeed, this is the asymptotic behavior of $\dot{g}_{ij}$ that is consistent with the one of $\pi^{ij}$ dictated by (\ref{asymptotics}). In particular, these conditions typically lead to finite functionals of the field variables, as, for example, the kinetic term $\int d^3x \pi^{ij} \dot{g}_{ij}$.

The object
\begin{equation}
 a_i = \partial_i \ln N
\end{equation}
transforms as a vector under the spatial sector of the foliation-preserving diffeomorphisms and as a scalar under the time transformations. That is, under $\mbox{Diff}(\mathcal{M},F)$ transformations defined by $\delta t = f(t)$ and $\delta x^i = \zeta^i (t,\vec{x})$, $a_i$ transforms according to
\begin{equation}
 \delta a_i = 
 \zeta^j \partial_j a_i + \partial_i \zeta^j a_j + f \dot{a}_i \,.
\end{equation}
Therefore, it can be consistently used \cite{Blas:2009qj} to add terms to the original Lagrangian of the Ho\v{r}ava theory.

The second-order action (without a cosmological constant), which is the most general one preserving time and spatial parity, is given by 
\begin{equation}
 S = 
 \int dt d^3 x \sqrt{g} N ( G^{ijkl} K_{ij} K_{kl} + R 
 + \alpha a_i a^i + \beta \nabla_i a^i ) \,,
\end{equation}
where $\alpha$ and $\beta$ are arbitrary coupling constants and
\begin{eqnarray}
K_{ij} & = & \frac{1}{2N} ( \dot{g}_{ij} - 2 \nabla_{(i} N_{j)} ) \,,
\\[1ex]
G^{ijkl} & = &
\frac{1}{2} \left( g^{ik} g^{jl} + g^{il} g^{jk} \right) 
- \lambda g^{ij} g^{kl} \,.
\end{eqnarray}
For the case of $\lambda \neq {1}/{3}$ the inverse of $G^{ijkl}$ is given by
\begin{equation}
\mathcal{G}_{ijkl} = 
\frac{1}{2} (g_{ik} g_{jl} + g_{il} g_{jk} )
- \frac{\lambda}{3\lambda - 1} g_{ij} g_{kl} \,.
\label{inverseg}
\end{equation}

Asymptotically (\ref{asymptotics}), $\sqrt{g} N \nabla_i a^i$ is of order $\mathcal{O}(r^{-3})$ whereas $\sqrt{g} N a_i a^i$ is $\mathcal{O}(r^{-4})$. Therefore, although one can express locally the former as the negative of the latter plus a spatial divergence, their contributions to the potential are definitely different due to the finite contribution of $\sqrt{g} N \nabla_i a^i$ at the spatial infinity. After the integration by parts, we get the final form of the Lagrangian,
\begin{equation}
 S = 
 \int dt \left[ \int d^3 x \sqrt{g} N ( 
 G^{ijkl} K_{ij} K_{kl} + R + \tilde{\alpha} a_i a^i )
 + \beta \Phi_N \right] \,,
 \label{action}
\end{equation}
where
\begin{equation}
\Phi_N \equiv
 \oint\limits_\infty d\Sigma_i \partial_i N \,,
\hspace{2em}
 \tilde{\alpha} = \alpha - \beta \,.
\end{equation}
To get this form of $\Phi_N$ we have used the asymptotics (\ref{asymptotics}). Since the flux $\Phi_N$ is a surface term at spatial infinity, its functional derivatives vanish provided the variations have an asymptotic behavior of $\mathcal{O}(r^{-2})$, the one consistent with the asymptotic assumptions in (\ref{asymptotics}) and on $\dot{g}_{ij}$ and $\dot{N}$. Consequently, there are no contributions of $\Phi_N$ to the field equations or to the Poisson brackets. In fact, Poisson brackets are distributions acting on functions of compact support; hence the Dirac delta and any of its derivatives acting on such functions at the boundary give a vanishing contribution.

We obtain the Hamiltonian from the action (\ref{action}). The momentum $\pi^{ij}$ has the universal form
\begin{equation}
\frac{\pi^{ij}}{\sqrt{g}} = 
G^{ijkl} K_{kl} \,.
\end{equation}
This implies that the velocities $\dot{g}_{ij}$ can be completely solved in favor of $\pi^{ij}$ if the condition $\lambda \neq 1/3$ is verified. We assume this condition throughout this paper. After performing the Legendre transformation and adding the primary constraint $\phi$, we get the Hamiltonian
\begin{equation}
 \begin{array}{rcl}
 H & = &
 {\displaystyle \int d^3x ( N \tilde{\mathcal{H}} + N_i \tilde{\mathcal{H}}^i 
 + \sigma \phi )  - \beta \Phi_N \,, }
 \\[2ex]
 \tilde{\mathcal{H}} & \equiv &
 \mathcal{G}_{ijkl} {\displaystyle\frac{\pi^{ij} \pi^{kl}}{\sqrt{g}}} 
 - \sqrt{g} ( R + \tilde{\alpha} a_i a^i ) \,,
 \\[2ex]
 \tilde{\mathcal{H}}^i & \equiv &
 - 2 \nabla_j \pi^{ji} \,,
 \end{array}
\label{prehamiltonian}
\end{equation}
where $\mathcal{G}_{ijkl}$ is the inverse of $G^{ijkl}$ and $\sigma$ is a Lagrange multiplier that transforms as a scalar under spatial coordinate transformations.

The Legendre transformation automatically incorporates the primary constraint $\tilde{\mathcal{H}}^i$ into the Hamiltonian. However, notice that $\tilde{\mathcal{H}}^i$ generates spatial coordinate transformations only in $g_{ij}$ and $\pi^{ij}$. In order to carry out a clear treatment, we would like to compute brackets with the full generator, which must include also the generator of the spatial coordinate transformations on $N$ and $\phi$ (in any case, we can always go to the gauge $N_i = 0$ and forget about these generators, but we prefer to perform a general treatment). The generator of infinitesimal spatial coordinate transformations on $N$ and $\phi$ is $\phi \partial_i N$, which vanishes in the constrained surface. Therefore, we redefine the momentum constraint by
\begin{equation}
 \mathcal{H}^i = \tilde{\mathcal{H}}^i + \phi \partial^i N 
\label{momentumconstraint}
\end{equation}
and replace $\tilde{\mathcal{H}}^i$ by $\mathcal{H}^i$ in the Hamiltonian (\ref{prehamiltonian}). A similar consideration was included in Ref.~\cite{Kluson:2010nf}.

To ensure the preservation in time of the $\phi$ constraint we need to compute only its Poisson bracket with $\int d^3x N\tilde{\mathcal{H}}$. We obtain
\begin{equation}
 \{ \phi , H \} = 
 - \tilde{\mathcal{H}} - 2 \tilde{\alpha} \sqrt{g} ( \nabla_i a^i + a_i a^i ) \,.
\end{equation}
Therefore, the density
\begin{equation}
 \mathcal{H} \equiv
 \tilde{\mathcal{H}} + 2 \tilde{\alpha} \sqrt{g} ( \nabla_i a^i + a_i a^i )
\label{hamiltonianconstraint}
\end{equation}
is a secondary constraint of the theory. Again, the expression $\sqrt{g} N ( \nabla_i a^i + a_i a^i )$ is an exact divergence whose integral gives the flux of $N$ at infinity. Thus, by solving $\tilde{\mathcal{H}}$ in terms of $\mathcal{H}$ and replacing it into the Hamiltonian, which adds another boundary term proportional to $\Phi_N$, we may write the Hamiltonian as a sum of constraints and a boundary term. After doing this we arrive at the final form of the Hamiltonian	
\begin{eqnarray}
 H & = & 
 \int d^3x ( N \mathcal{H} + N_i \mathcal{H}^i + \sigma \phi ) 
 - \tilde{\beta} \Phi_N\,,
\label{hamiltonian}
\\
 \mathcal{H} & = &
 \mathcal{G}_{ijkl} \frac{\pi^{ij} \pi^{kl}}{\sqrt{g}} 
 + \sqrt{g} ( - R  +  \tilde{\alpha} ( 2 \nabla_i a^i  + a_i a^i ) )
\label{H} \,,
\\
 \tilde{\beta} & = & 2 \alpha - \beta  \,,
\end{eqnarray}
and $\mathcal{H}^i$ is given in (\ref{momentumconstraint}).

If we use variations in the strong sense as in \cite{Regge:1974zd}, Hamiltonian (\ref{hamiltonian}) is functionally differentiable if and only if $\tilde{\beta} = 2 \alpha$, that is, $\beta =  0$. The remaining boundary term is then relevant in this mathematical sense. From the physical point of view, it contributes to the gravitational mass of the theory. Thus we see that, under the variational scheme of Ref.~\cite{Regge:1974zd}, the $\nabla_i a^i$ term must be excluded from the Lagrangian. However, we notice that under the asymptotic conditions we are considering, where $\dot{g}_{ij}$ and $\dot{N}$ behave as $\mathcal{O}(r^{-2})$, the arbitrary variations $\delta g_{ij}$ must be of the same order $\mathcal{O}(r^{-2})$. Otherwise the integrability of the Lagrangian is violated. Assuming then this behavior, it turns out that the variation of the action gives rise to a boundary integral in terms of the variations which become zero. The functional differentiability of the action is then satisfied without adding boundary terms \emph{a la} Regge and Teitelboim. In any case, we think it would be the quantum theory which will determine if there are quantum contributions or not to the boundary term. The coefficient beta will then be determined from quantum corrections. An important point related to it is the quantum stability of the action, a problem directly related to the positiveness of the gravitational mass. This problem has been recently addressed in Ref.~\cite{arXiv:1108.1835}. The quantum stability of the theory will determine bounds to the coefficient beta.

Condition $\mathcal{H}=0$ is a nonlinear elliptic PDE for $N$; hence in principle we can solve it for $N$. Indeed, for the variable $\sqrt{N}$ this condition becomes a linear, homogeneous, elliptic PDE,
\begin{equation}
\left( 4 \tilde{\alpha} \nabla^2 
- R +  \mathcal{G}_{ijkl} \frac{\pi^{ij} \pi^{kl}}{g} 
\right) \sqrt{N} = 0 \,.
\label{sqrtn}
\end{equation}
To check the compatibility with the boundary conditions, at the asymptotic limit we expand $\sqrt{N} = 1 + n$ and cast Eq.~(\ref{sqrtn}) as an equation for $n$. We get that the dominant terms are
\begin{equation}
4 \tilde{\alpha} \partial_i \partial_i n = R \,.
\label{asymptoticeqn}
\end{equation}
Since $R$ is of $\mathcal{O}(r^{-3})$, this equation dictates $n = \mathcal{O}(r^{-1})$, which is the assumed behavior in (\ref{asymptotics}).

It is worth studying a bit more the existence and uniqueness properties of Eq.~(\ref{sqrtn}), since $N$ is a dynamical variable of the theory which we expect to be fixed by the Hamiltonian constraint. Any source of indetermination on $N$ could lead to inconsistencies of the theory, or may force us to reinterpret the Hamiltonian constraint as a condition for another variable. The existence and uniqueness of the solution of Eq.~(\ref{sqrtn}) are rather nontrivial aspects because of the presence of the $R$ and $\pi^2$ terms in the operator. We may give a partial answer to this question by linearizing the equation for all field variables, which can be achieved by a perturbative analysis around a Minkowski background. We expand $\sqrt{N} = 1 + n$, $g_{ij} = \delta_{ij} + h_{ij}$ and $\pi^{ij}$ is considered of first order in perturbation. Under these settings, a perturbative expansion around Minkowski space-time is equivalent to going to the asymptotic limit. Hence we may read the linearized version of Eq.~(\ref{sqrtn}) easily from Eq.~(\ref{asymptoticeqn}), obtaining
\begin{equation}
4 \tilde{\alpha} \partial_i \partial_i n = 
\partial_i \partial_j h_{ij}  - \partial_i \partial_i h\,.
\label{lineareqn}
\end{equation}
This is an ordinary Poisson equation in flat space, subject to the Dirichlet boundary condition $n |_\infty = 0$ and with asymptotic decay of the source of $\mathcal{O}(r^{-3})$, which is fast enough to ensure the finiteness of the solution. Therefore, the solution of Eq.~(\ref{lineareqn}) with the given boundary condition exists and is unique.

Moreover, since the differential operator is elliptic, we can give a result on nonperturbative grounds about the existence and uniqueness of the solution of the PDE (\ref{sqrtn}). If the nonderivative terms of Eq.~(\ref{sqrtn}) satisfy the condition
\begin{equation}
 \mathrm{sgn}(\tilde{\alpha}) \left(R -  g^{-1} \mathcal{G}_{ijkl} \pi^{ij} \pi^{kl} \right) 
 \geq 0 
 \label{condition}
\end{equation}
in all the spatial submanifold, then the Lax-Milgram theorem can be used to obtain directly that the \emph{weak} solution of (\ref{sqrtn}) subject to the prescribed boundary condition exists and is unique\footnote{With weak we mean a solution in the sense of distributions: if $L$ stands for the operator acting on $\sqrt{N}$, then there exists a unique function $\sqrt{N}$ that satisfies the boundary conditions and $\int d^3x f L \sqrt{N} = 0$ for any smooth function $f$ of compact support.}. It would be nice to prove the existence and uniqueness of the weak solution when the condition (\ref{condition}) is relaxed.

From the previous experience \cite{Bellorin:2010je} with the model without the $a_i$ terms, one expects that $\phi$ and $\mathcal{H}$ are of second-class behavior. Hence it is useful to have the Poisson brackets between them. The nonzero brackets, evaluated on the constrained phase space, are equal to
\begin{eqnarray}
\{ \int d^3 x \epsilon \mathcal{H} , \int d^3y \eta \mathcal{H} \} 
 & = & 
 \int d^3x \left( \frac{2(\lambda - 1)}{3\lambda - 1} 
 ( \eta \nabla^2 \epsilon ) \pi 
 + 2 \tilde{\alpha} 
   (\eta \partial^i \epsilon ) \mathcal{G}_{ijkl} a^j \pi^{kl} \right)
- (\epsilon \leftrightarrow \eta)  \,,
\nonumber \\
\label{brackethh}
\\
 \{ \int d^3x \epsilon \mathcal{H} , \int d^3y \eta \phi  \} & = & 
 2 \tilde{\alpha} \int d^3x \sqrt{g} \eta ( \nabla^2 \epsilon - \nabla_i (\epsilon a^i) ) N^{-1} \,.
\label{brackethphi}
\end{eqnarray}

Now we demand the time preservation of $\mathcal{H}$. This requires computing its Poisson brackets with $\int d^3x N\mathcal{H}$ and $\int d^3x \sigma \phi$, which can easily be read from (\ref{brackethh}) and (\ref{brackethphi}). We obtain
\begin{equation}
\begin{array}{rcl}
 \{ \mathcal{H}, H \} & = &
 2 \tilde{\alpha} \sqrt{g} 
   [ \nabla^2 (\sigma / N) + a^i \partial_i (\sigma / N) ]
 - 2 \tilde{\alpha} N \mathcal{G}_{ijkl} [ \nabla^i ( a^j \pi^{kl} )
    + 2 a^i a^j \pi^{kl} ] 
\\[1.5ex]
&& + {\displaystyle\frac{2(\lambda - 1)}{3\lambda -1}} N^{-1} \nabla_i (N^2 \nabla^i \pi) \,.
\end{array}
\end{equation}
Therefore, the preservation in time of $\mathcal{H}$ implies the following equation for $\sigma$
\begin{eqnarray}
&& \tilde{\alpha} [ \nabla^2 (\sigma / N) + a^i \partial_i (\sigma / N) ] =
 \mathcal{J} \,,
\label{sigmaeq}
\\[1ex]	
&& \mathcal{J} \equiv
 - \left(\frac{\lambda - 1}{3\lambda -1} \right)
    \frac{N^{-1}}{\sqrt{g}} \nabla_i (N^2 \nabla^i \pi)
 + \tilde{\alpha} \frac{N}{\sqrt{g}} \mathcal{G}_{ijkl} [ \nabla^i ( a^j \pi^{kl} )
    + 2 a^i a^j \pi^{kl} ]   \,.
\end{eqnarray}

Equation (\ref{sigmaeq}) is a second-order, linear, elliptic PDE for $\sigma$ with a source term independent of $\sigma$. Thus we can solve this equation for $\sigma$ in a unique way if the operator is compatible with the prescribed boundary conditions. First, by varying the canonical action with respect to $\phi$ we get the relation $\dot{N} = \sigma$. From this relation we may read the asymptotic behavior of $\sigma$, but being careful with the subtleties arising with the time dependence. We recall that in (\ref{asymptotics}) it is assumed that $\dot{g}_{ij}$ asymptotically is of $\mathcal{O}(r^{-2})$. By going back to Eq.~(\ref{asymptoticeqn}), we note that $\dot{n}$ is of $\mathcal{O}(r^{-2})$. Therefore, we conclude that $\sigma = \mathcal{O}(r^{-2})$. Now we analyze the asymptotic behavior of Eq.~(\ref{sigmaeq}). In the source $\mathcal{J}$  the asymptotically leading term is the first term, which decays as $\mathcal{O}(r^{-4})$. The dominant term for $\sigma$ in the left-hand side is $\propto\partial_i \partial_i \sigma$; hence we get
\begin{equation}
\tilde{\alpha} \partial_i \partial_i \sigma = \mathcal{O}(r^{-4})  \,,
\end{equation}
which is satisfied by $\sigma = \mathcal{O}(r^{-2})$. Since we can solve Eq.~(\ref{sigmaeq}) for the Lagrange multiplier $\sigma$, Dirac's algorithm for the constraints ends at this step.

We have ended up with the momentum constraint $\mathcal{H}^i$ and the constraints $\phi$ and $\mathcal{H}$. From (\ref{brackethh}) and (\ref{brackethphi}) we confirm that $\phi$ and $\mathcal{H}$ are the second-class constraints of the theory. The physical degrees of freedom are given by
\begin{equation}
\begin{array}{l}
\mbox{(\# Physical D.O.F.)} =  
\nonumber \\[1ex]
 \hspace*{2em}  \mbox{(\# Canonical var.)} 
   - 2 \times \mbox{(\# 1$^{\mbox{\tiny st}}$ class const.)} 
     -\mbox{(\# 2$^{\mbox{\tiny nd}}$ class const.)} 
     = 6 \,.
\end{array}
\end{equation}
This leaves us with six independent degrees of freedom in the canonical space, which are equivalent to three propagating even modes. Two of them correspond to the graviton and the remaining one is an even scalar mode. Such a scalar mode is activated in the metric $g_{ij}$, as was focused in Ref.~\cite{Blas:2009qj}.

By going to the $\tilde{\alpha} = 0$ limit we get that Eqs.~(\ref{sqrtn}) and (\ref{sigmaeq}) becomes, respectively,
\begin{eqnarray}
- R +  \mathcal{G}_{ijkl} \frac{\pi^{ij} \pi^{kl}}{g} 
 & = & 0 \,,
\label{hlr}
\\[1ex]
 (\lambda - 1) \nabla_i (N^2 \nabla^i \pi) & = & 0 \,.
\label{pi}
\end{eqnarray}
We see that $N$ decouples from (\ref{hlr}) and $\sigma$ from (\ref{pi}). The first of these equations defines the Hamiltonian constraint of the $\lambda R$ model \cite{Bellorin:2010je}. The only solution of the second equation that satisfies (\ref{asymptotics}) is $\pi= 0$, which is interpreted as a second-class constraint in this limit. The preservation in time of $\pi$ yields the second-class constraint $(\nabla^2 - R) N = 0$. Finally, the preservation of this last constraint yields an equation for $\sigma$ of the type $(\nabla^2 - R) \sigma = \tilde{\mathcal{J}}$. This limit is physically equivalent to general relativity \cite{Bellorin:2010je}.


\section{A conformal transformation depending on $N$}
One might be wondering about to what extent the spatial derivatives of $N$ arising in the action, Hamiltonian and constraints can be simplified by a conformal transformation of the metric depending on $N$. To focus this issue, we start by noting that the conformal transformation
\begin{equation}
\tilde{g}_{ij} = N^{2\epsilon} g_{ij} \,,
\hspace{2em}
\tilde{\pi}^{ij} = N^{-2\epsilon} \pi^{ij} \,,
\label{conformal}
\end{equation}
where $\epsilon$ is an arbitrary real parameter, is a canonical transformation if we compensate it with the transformation
\begin{equation}
\tilde{\phi} = \phi - 2 \epsilon N^{-1} \pi \,.
\label{p}
\end{equation}
Indeed, we get
\begin{equation}
 \pi^{ij} \dot{g}_{ij} + \phi \dot{N} =
 \tilde{\pi}^{ij} \dot{\tilde{g}}_{ij} + \tilde{\phi} \dot{N} \,,
\end{equation}
and the nonvanishing Poisson brackets are
\begin{equation}
\begin{array}{rcl}
 \{ \tilde{g}_{ij}(t,\vec{x}) , \tilde{\pi}^{kl}(t,\vec{y}) \} & = &
 \frac{1}{2}(\delta_i^k \delta_j^l + \delta_i^l \delta_j^k)
 \delta(\vec{x}-\vec{y}) \,,
 \\[2ex]
 \{ N (t,\vec{x}) , \tilde{\phi}(t,\vec{y}) \} & = &
 \delta(\vec{x}-\vec{y}) \,.
\end{array}
\label{newbrackets}
\end{equation}
These brackets were computed in terms of the old variables, but from now on they can be redefined in terms of the canonical variables $(\tilde{g}_{ij} , \tilde{\pi}^{ij})$ and $(N,\tilde{\phi})$, (\ref{newbrackets}) being the fundamental brackets. Now the primary constraint $\phi = 0$ takes the form
\begin{equation}
\tilde{\phi} = -2 \epsilon N^{-1} \tilde{\pi} \,.
\label{phitilde}
\end{equation}

After performing the transformations (\ref{conformal}) - (\ref{p}), we get that the term $\int d^3x N \tilde{\mathcal{H}}$ of the Hamiltonian given in (\ref{prehamiltonian}) becomes
\begin{equation}
\int d^3x N \tilde{\mathcal{H}} = 
\int d^3x \sqrt{\tilde{g}} N \left( 
 N^{3\epsilon} \tilde{\mathcal{G}}_{ijkl}
 \frac{\tilde{\pi}^{ij} \tilde{\pi}^{kl}}{\tilde{g}}
 - N^{-\epsilon} ( \tilde{R} + \chi \tilde{g}^{ij} a_i a_j )
    \right) 
- 4 \epsilon \Phi_N \,,
\label{confhamiltonian}
\end{equation}
where $\chi = \tilde{\alpha} + 2 \epsilon^2 - 4 \epsilon$. Thus, we can drop the term with derivatives of $N$ out of the bulk integral in (\ref{confhamiltonian}) by putting $\chi = 0$, which gives 
\begin{equation}
\epsilon = 1 \pm \sqrt{1 - \tilde{\alpha}/2}
\label{epsilon}
\end{equation}
under the condition 
\begin{equation}
\tilde{\alpha} \leq 2 \,.
\label{condalpha}
\end{equation}

Therefore, for the sector of the space of parameters where (\ref{condalpha}) is verified, the (bulk) Hamiltonian can be written without any explicit dependence on the derivatives of $N$. However, derivatives of $N$ still arise when one evaluates the condition for the preservation in time of the constraint (\ref{phitilde}), that is, the Hamiltonian constraint. Curiously, there is a special value of the coupling constant $\tilde{\alpha}$ for which the Hamiltonian gets an even simpler form. For $\tilde{\alpha} = 2$ the formula (\ref{epsilon}) yields $\epsilon = 1$ and (\ref{confhamiltonian}) takes the form
\begin{equation}
\int d^3x N \tilde{\mathcal{H}} = 
\int d^3x \sqrt{\tilde{g}} \left( 
 N^{4} \tilde{\mathcal{G}}_{ijkl}
 \frac{\tilde{\pi}^{ij} \tilde{\pi}^{kl}}{\tilde{g}}
 - \tilde{R}   \right) 
- 4 \Phi_N \,.
\end{equation}
We may see that in this case there is not dependence on $N$ of the potential. This implies in particular that the Hamiltonian constraint acquires a totally algebraic dependence on $N$. This constraint becomes
\begin{equation}
\tilde{\mathcal{G}}_{ijkl}
 \tilde{\pi}^{ij} \tilde{\pi}^{kl} N^4 =
 \tilde{g} \tilde{R} \,.
\end{equation}
It would be interesting to elucidate the role of special values of the coupling constants, such as $\tilde{\alpha} = 2$, from the point of view of the renormalization-group flow of the theory.


\section{Conclusions}
We have performed the Hamiltonian analysis to the lowest-order effective action of the complete nonprojectable Ho\v{r}ava theory. Following the proposal of Blas, Pujol\`as and Sibiryakov, this action contains a term proportional to $(\partial_i \ln N)^2$ and a boundary term which comes from the $\nabla^2 N$ term. This model has the momentum constraint as the first-class constraint. The second-class constraints are the vanishing of the momentum conjugated to $N$ and the constraint needed for its preservation in time, which is analogous to the so-called Hamiltonian constraint in general relativity. These constraints reduce to six the number of independent canonical variables, corresponding to three physical modes. It is remarkable that the extra scalar mode in this model is even, that is, evolves with a second-order equation. This fact was one of the motivations of Blas, Pujol\`as and Sibiryakov for introducing the extra terms in Ref.~\cite{Blas:2009qj}. This even scalar must be contrasted with the odd extra scalar found in Refs.~\cite{Blas:2009yd,Bellorin:2010te} for models with quadratic-curvature terms in the potential and without $a_i$-dependent terms. 

When compared with the lowest-order effective action of the original Ho\v{r}ava theory, which is the $\lambda R$ model studied by us in Ref.~\cite{Bellorin:2010je}, the model studied in this paper lacks two second-class constraints. From the Hamiltonian-analysis point of view, this is the origin for the extra even scalar mode absent in the $\lambda R$ model, which is physically equivalent to general relativity.

The Hamiltonian constraint is a second-order elliptic PDE for $N$ totally compatible with the standard (flat) asymptotic behavior of all the gravitational variables. In order to give support for the strategy of solving for $N$ this constraint, we have linearized it and have found that it becomes a Poisson equation that can be solved in a closed way for $N$. Moreover, if the nonderivative terms satisfy a kind of non-negativity condition, the standard theory of PDEs ensures that the solution for $N$ exists and is unique, at least in the sense of distributions. This is a totally nonperturbative result. It will be interesting to give the general proof of the existence and uniqueness of the solution for $N$. The preservation in time of the Hamiltonian constraint yields a second-order elliptic PDE for the Lagrange multiplier $\sigma$ of the theory compatible with the asymptotic conditions, hence ending Dirac's algorithm at this step. In this equation the operator acting on $\sigma$ does not contain nonderivative terms, which makes it simpler to conclude that the solution exists and is unique.

We have also shown that for a particular value of one of the coupling constants the Hamiltonian constraint acquires an algebraic dependence on $N$ after a suitable canonical transformation.

Upon these results we conclude that the lowest-order truncation of the nonprojectable Ho\v{r}ava theory with the terms of Blas, Pujol\`as and Sibiryakov has a consistent and closed algebra of constraints, as was indicated in Ref.~\cite{Kluson:2010nf}, but the theory manifestly deviates from Einstein's general relativity at the deep IR, unlike the original Ho\v{r}ava theory \cite{Bellorin:2010je}. For the viability of the theory beyond the mathematical consistency, the physics of the extra scalar mode should be contrasted with the phenomenology, as was concluded by Blas, Pujol\`as and Sibiryakov in Ref.~\cite{Blas:2009qj}.

The consistent structure we have found for the Hamiltonian formulation of the effective theory depends crucially on the form of the operators acting on $N$ in the Hamiltonian constraint and on $\sigma$ in its corresponding equation. We have seen that these operators are of second-order elliptic form. However, from our analysis it cannot be asserted that the Hamiltonian formulation will remain consistent when higher-order operators are included, particularly the $z=3$ terms required for renormalizability, since they raise the order of the mentioned operators. For example, this will happen in the Hamiltonian constraint when the fourth-order term $(a_i a^i)^2$ is included. A promising feature about this point is that, if spatial parity is imposed as preserved symmetry, odd-dimensional operators are forbidden \cite{Blas:2009qj}. Thus, the elliptic form of the equations for $N$ and $\sigma$ possibly will be preserved by the higher-order operators.

The model we have studied in this paper can be extended in several ways. An immediate step would be the inclusion of a cosmological constant in the Lagrangian. In this case the Hamiltonian analysis should be started from the very beginning since the boundary conditions and the asymptotic behavior of the field variables must be modified. We do not expect any modification in the \emph{number} of constraints that emerge from Dirac's procedure, since the Hamiltonian constraint will depend on $N$ and its preservation in time will generate an equation for $\sigma$. Moreover, this equation will remain intact since a cosmological-constant term has no derivatives, hence it does not contribute to the source $\mathcal{J}$ of $\sigma$. However, the details about the compatibility between these equations and the boundary conditions should be investigated carefully in order to arrive at a definitive conclusion about the consistency of the Hamiltonian formulation of the theory with cosmological constant.

\begin{center}
{* \hspace*{5em} * \hspace*{5em} *}
\end{center}

\vspace*{1em}
\noindent {\bf Note added}: 
While we were preparing the first version of our manuscript, the paper ``Hamiltonian Structure of Ho\v{r}ava Gravity'', by W. Donnelly and T. Jacobson \cite{Donnelly:2011df}, was published on the Web. With regard to the second-order action, that paper contains the Hamiltonian and the constraints, and we agree with their results. They also argued that the preservation in time of the Hamiltonian constraint leads to a condition for the Lagrange multiplier $\sigma$ (called $Nw$ by them). We have found explicitly this condition in Eq.~(\ref{sigmaeq}). In addition, they found that for the asymptotically flat case the Hamiltonian can be written as a sum of constraints plus a boundary contribution. We also found this structure for the Hamiltonian, but also the contribution of the $\nabla_i a^i$ term at the boundary. However, we must stress that, if variations in the strong sense of Ref.~\cite{Regge:1974zd} are used, then the $\nabla_i a^i$ term cannot be included in the Lagrangian in order to obtain a functionally differentiable action with those variations.


\end{document}